# Mist Flow Visualization for Round Jets in Aerosol Jet® Printing

James Q. Feng

Optomec, Inc., 2575 University Avenue, #135, St. Paul, MN 55114, USA

jfeng@optomec.com

**Abstract**

With the microdroplets of water serving as light scattering particles, the mist flow patterns of round micro-jets can be visualized using the Aerosol Jet® direct-write system. The visualization images show that the laminar mist jet (with sheath-to-mist ratio $Y = 1:1$) appears to extend to more than 20 times the diameter of nozzle orifice $D$ for jet Reynolds number $Re < 600$, especially with $D = 0.3$ mm and less. For smaller jets (e.g., with $D = 0.15$ mm), laminar collimated mist flow might be retained to 40x$D$ for $Re < 600$ and for $Re \sim 1500$ within 20x$D$ from the nozzle. The laminar part of mist flow associated with larger jets (e.g., with $D = 1.0$ mm for $Re < 600$) tends to exhibit noticeable gradual widening due to viscous diffusion. For free jets, their breakdown length—the distance from nozzle where transition from laminar to turbulent mist flow takes place as signaled by a rapid widening of mist stream—is shown to decrease with increasing $Re$. The presence of impingement wall tends to prevent turbulence development, even when the wall is placed further downstream of the free-jet breakdown length for a given $Re$. The critical $Re$ for an impinging jet to develop turbulence increases as the standoff $S$ is reduced. The mist flow of impinging jet of $D = 1.0$ mm seems to remain laminar even for $Re > 4000$ at $S = 12$ mm.

## 1 Introduction

As an effective additive manufacturing tool, the Aerosol Jet® direct-write technology of Optomec enables fabrication of microscale functional devices for various applications (cf. Zollmer et al., 2006; Hedges et al., 2007; Kahn, 2007; Christenson et al., 2011; Paulsen et al., 2012; Renn et al., 2017). It deposits functional ink materials in a form of high-speed mist stream with an impinging jet flow, based on the mechanism of inertial impaction of microdroplets of diameters typically ranging from 1 to 5 microns (Renn, 2006; Binder et al., 2014; Feng, 2015). With an appropriately adjusted flow rate according to the diameter of deposition nozzle orifice (ranging from 0.1 to 1 mm), the mist stream appears to be substantially collimated such that well-controlled print quality can usually be achieved with a standoff of several millimeters between the deposition nozzle and substrate, easily allowing the Aerosol Jet® system to deposit ink materials onto three-dimensional substrates of complicated geometries.

In terms of fluid dynamics, jet flows have been extensively studied by numerous authors. Because of the inflection point in its velocity profile, a free submerged jet is unstable based on inviscid parallel flow theory known as the Rayleigh-Fjortoft criterion (cf. Drazin and Reid, 1981). Photographs of free submerged jets often show a laminar "potential core" emerging from the nozzle, with growing waves downstream rolling up into ring vortices along the edges due to Kelvin-Helmholtz instability driven by the local velocity gradients (Todde et al., 2009). The spatial distances for evolution from the nozzle exit, to rolling up of the vortex layer, to coalescence of ring vortex pairs, and eventually to the development of turbulent eddies, were found to shorten with increasing Reynolds number (Becker and Massaro, 1968; Kwon and Seo, 2005; Abdel-Rahman, 2010). However, the visualization images of 2D jets from microslot nozzles with width less than 0.2 mm did not exhibit the vortex forming and merging processes like those typically seen in previously studied larger-scale jets with diameters in the range of centimeters (Gau et al., 2009). The breakdown length--distance (in units of the slot width) for transition from laminar to turbulent flow with those 2D microjets--seems to be much longer than that with the macrojets, though also monotonically decreasing with the jet Reynolds number.

In the presence of an impingement wall, the impinging jets may remain laminar when reaching the wall especially when the standoff is less than the free-jet breakdown length. For consistent Aerosol Jet® printing, the process of ink deposition should be associated with a steady impinging jet having laminar mist flow for a standoff often greater than ten times the nozzle orifice diameter. Yet most research articles in the literature have been focusing on turbulent impinging jets at large Reynolds numbers, for effective heat and mass transfer in various industrial applications (Popiel and Trass, 1991; Zuckerman and Lior, 2006). A study of impinging laminar jets at moderate Reynolds numbers (Bergthorson et al., 2005) was carried out only with a standoff less than 1.5 times the nozzle diameter of 0.99 cm, as relevant to the authors' interest in strained flames. The present work is devoted to a systematic investigation of the images from mist flow visualization from microscale round nozzles, to provide valuable practical guidance for the Aerosol Jet® printing, and to gain general insights into the fundamental behavior of micro-jets less studied by previous authors.

## 2 Experimental Apparatus

Jet flow visualizations have been commonly performed by adding small particles, such as smoke or microspheres, to the flowing fluid such that scattered illumination light can make the flow pattern visible. In the present work, such small particles are provided naturally by the Aerosol Jet® direct-write system with its ink mist generated from an atomizer. For simplicity, the ink used here is just the deionized (DI) water, and the small particles are water microdroplets in the diameter range of 1 to 5 microns when arriving at the deposition nozzle, out of which the flow visualization is taking place. With those mist microdroplets being illuminated, a gas jet flow field under investigation becomes visible.

2.1 Aerosol Jet® System

As shown in Fig. 1, the Aerosol Jet® direct-write system consists of an atomizer, a deposition head, a mist transport-conditioning channel between atomizer and deposition head, and a substrate holding stage with adequately accurate motion control as well as a mist flow switching device. The atomizer generates a mist of microdroplets of the functional ink, with diameters becoming 1 to 5 microns through the transport-conditioning channel when arriving the deposition head, where the mist is wrapped with a co-flowing sheath gas and aerodynamically focused through the converging channel in the deposition nozzle to form a collimated high-speed mist jet for effective inertial impaction onto the substrate. With a standoff of several millimeters, noncontact printing of various patterns on substrates of complex geometries is enabled with CAD-driven relative motions between the substrate and deposition head.

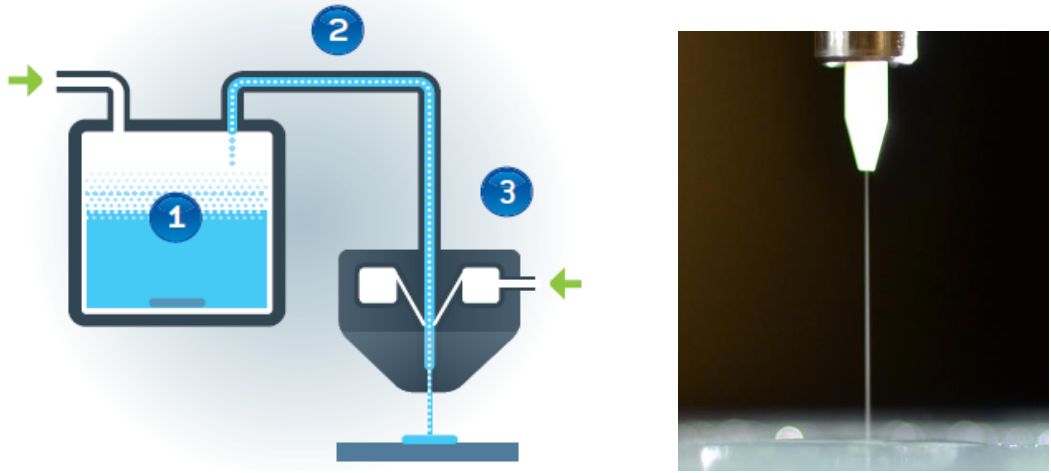

Fig. 1  Schematics of the Aerosol Jet® direct-write system: (1) Atomizer that generates mist of microdroplets of functional inks; (2) Mist transport-conditioning channel that delivers a concentrated mist of ink microdroplets; (3) Deposition head to form high-speed collimated mist stream through an aerodynamic focusing nozzle with sheath gas (with arrows indicate the mist carrier gas inlet and nozzle sheath gas inlet).   The photography on right shows visualized mist stream from the deposition nozzle.

Inside the nozzle there is a tapered converging channel (with a taper half angle of a few degrees), to gradually accelerate the mist flow while being aerodynamically focused. With the co-flowing sheath gas, the mist stream is squeezed toward the channel center, becoming narrower than the channel size for printing features much finer than the diameter of nozzle orifice. The printed feature size can be controlled by adjusting the ratio of sheath gas and mist flow rates. Another functionality of sheath gas is to prevent direct contact between mist droplets and nozzle channel wall so as to avoid clogging of small nozzle orifice by ink materials.

For mist jet flow visualization, a commercially available AJ200 system is used with a complete mist deposition system and a 2D motion stage including manually adjustable standoff between the deposition nozzle and substrate. With DI water placed in the atomizer, a visible mist stream of water microdroplets emerges from the tip of deposition nozzle displaying the jet flow pattern under adequate illumination (as shown by the photograph in Fig. 1).

2.2 Mist Flow Visualization Setup

As direct-write equipment, the Aerosol Jet® system (such as AJ200) comes with a set of mass flow controllers (MFCs) by ALICAT, to accurately control the mass flow rate within 1% of the MFC set value. Here the volumetric flow rate $Q$ of compressible gas is measured in units of "standard cubic centimeters per minute" (sccm), such that the mass flow rate remains constant while the actual volumetric flow rate of nitrogen varies with the local temperature and pressure. Thus, the value of jet mass flux $\rho U = 4 \rho_s Q / (\pi D^2)$ is a constant for given $Q$ and nozzle orifice diameter $D$, with $\rho$ and $\rho_s$ denoting the actual density of gas and that under standard conditions at $T_s = 273$ K and $P_s = 10^5$ Pa, e.g., $\rho_s = P_s / (R T_s) = 1.276$ kg/m$^3$ for dry nitrogen (which is typically used as the inert processing gas in Aerosol Jet® systems). The value of the jet Reynolds number is calculated according to

$$Re = \rho U D / \mu = 1.464 Q / D , \qquad (1)$$

with $Q$ in units of sccm and $D$ in millimeters assuming the dynamic viscosity of dry nitrogen $\mu = 1.85 \times 10^{-5}$ kg m$^{-1}$ s$^{-1}$ (at $T = 300$ K). Hence $Re = 732$ with $Q = 150$ sccm and $D = 0.3$ mm, while $Re = 586$ for $Q = 60$ sccm and $D = 0.15$ mm.

In the present work, deposition nozzles of various diameters $D$ are used with various nitrogen gas flow rate $Q$ so as to obtain a large range of jet Reynolds number $Re$ values relevant to Aerosol Jet® direct-write applications. It should be noted that the gas flow rate $Q$ here includes that of the sheath gas and the mist. The sheath-to-mist ratio $Y$ is nominally 1:1 but can be varied depending on specific applications with Aerosol Jet® printing. The standoff between the nozzle tip and substrate $S$ (in units of millimeter) is also varied to examine its effects on impinging jet behavior. When $S$ is set much larger than 60 times $D$, the flow is considered as that of a free-jet in an open space.

An AmScope LED Spot Light is used for illumination. A Nikon D3100 digital camera having a 14.2 megapixel CMOS sensor is used with a Nikon 105mm f/2.8D AF Micro-Nikkor lens, for taking the mist flow visualization images. The mist flow with DI water microdroplets becomes visible when the mist flow rate is at 20 sccm or greater.

**3 Results and Discussion**

The nozzles utilized in Aerosol Jet® printing can have a wide range of orifice diameters, $D$. The results presented here start with the case for a medium size nozzle of $D = 0.3$ mm in subsection 3.1, and followed by those of $D = 0.15$ mm and $D = 1.0$ mm in 3.2 and 3.3, respectively.

3.1 Cases of $D = 0.3$ mm

With free jets, Fig. 2 shows that the mist flow at $Re = 293$ remains laminar for extensive length (e.g., > 18 mm covered in the viewing area, gauged by the 4 mm long white ceramic nozzle tip outside the stainless steel holder of 4 mm diameter). The jet of $Re = 293$ seems to gradually widen due to viscous diffusion of momentum but no sign for turbulence within 18 mm of the viewing area. The mist flow of $Re = 586$ remains collimated up to 13 mm from the nozzle; then it breaks down to develop turbulent eddies downstream due to the mechanism of intrinsic instability, signaled by rapid widening of flow pattern. The transition from laminar to turbulent mist flow (sometimes called jet "breakdown") is clearly shown in the jet at $Re = 586$ and 1171, incepting at about 12 and 5 mm from the nozzle tip (which suggest likely breakdown lengths of 40 and 17, in units of $D$).

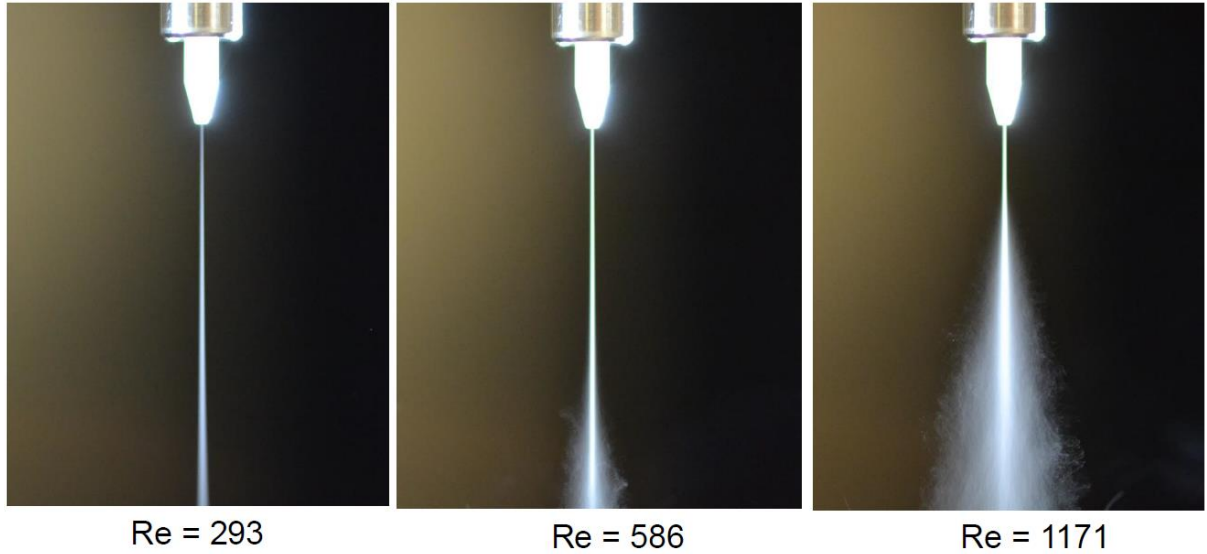

Fig. 2  Free-jet mist flows of $D = 0.3$ mm at Re = 293, 586, and 1171 (for $Q = 60$, 120, and 240 sccm) with sheath-to-mist ratio $Y = 1:1$.  The length of the white ceramic nozzle tip outside the stainless steel holder is 4 mm, which can effectively serve as a scale reference.

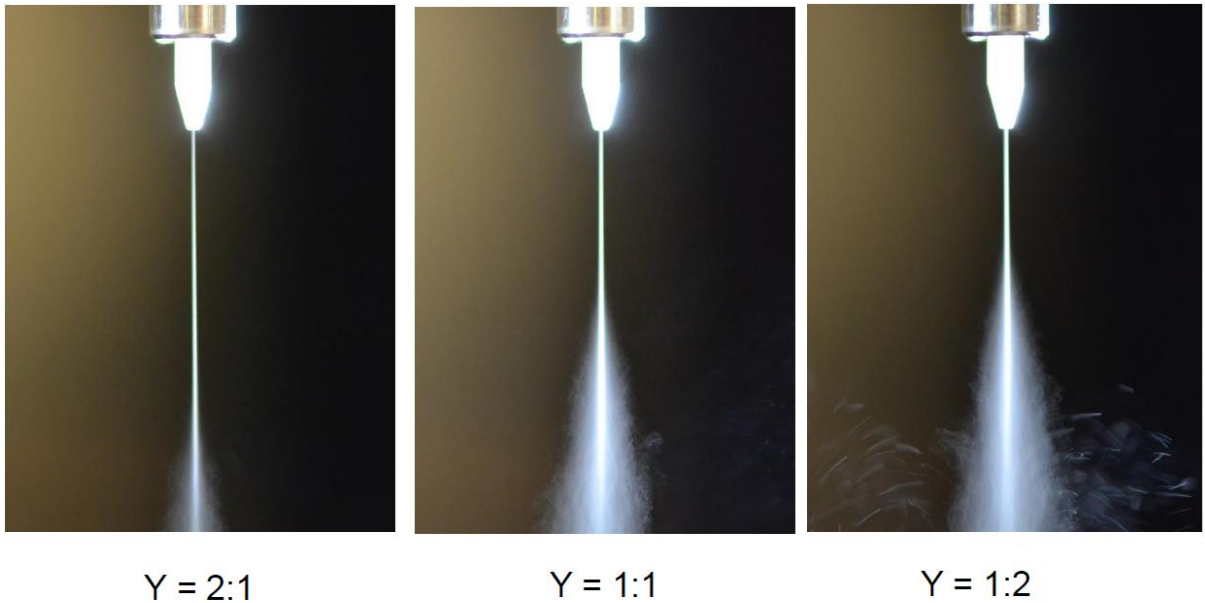

Fig. 3  As Fig. 2 but at $Re = 732$ (for $Q = 150$ sccm) with sheath-to-mist ratio $Y = 2:1$, 1:1, and 1:2 having corresponding breakdown length of 47, 27, and 20, respectively.

If the unstable waves are growing from the jet edges to form rolling up vortices and the develop into turbulent eddies (as seen in the smoke-wire visualization images of Popiel and Trass, 1991), increasing the sheath-to-mist ratio can effectively keep mist stream within the laminar potential core for a longer distance, away from the unstable waves along the jet edges. Therefore, increasing the sheath-to-mist ratio $Y$ can extend the breakdown length as shown in Fig. 3, enable deposition of ink materials at extended standoff with Aerosol Jet® printing for additive manufacturing. The images in Fig. 3 indicate the dependence of breakdown length upon sheath-to-mist ratio, reflecting the penetration rate of growing unstable waves along the jet edges. Hence, the Aerosol Jet® system, with its capability of easily adjusting the sheath-to-mist ratio, can also become an interesting research instrument for jet flow pattern investigation by "peeling the onion".

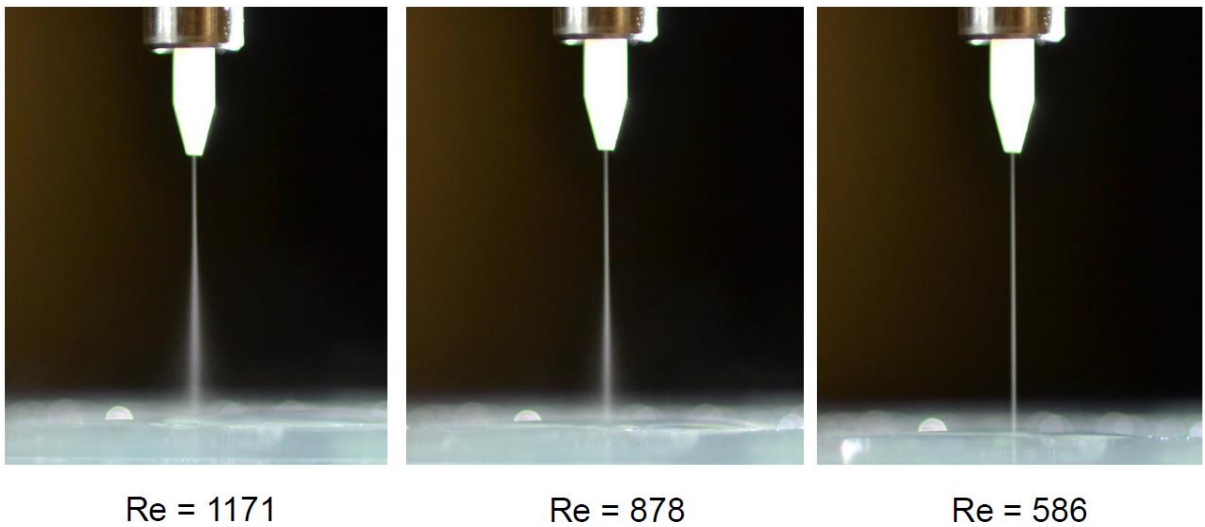

Fig.4 Impinging-jet mist flows of $D = 0.3$ mm at $Re = 1171$, 878, and 586 (for $Q = 240$, 180, and 120 sccm) with sheath-to-mist ratio $Y = 1:1$, for $S = 10$ mm.

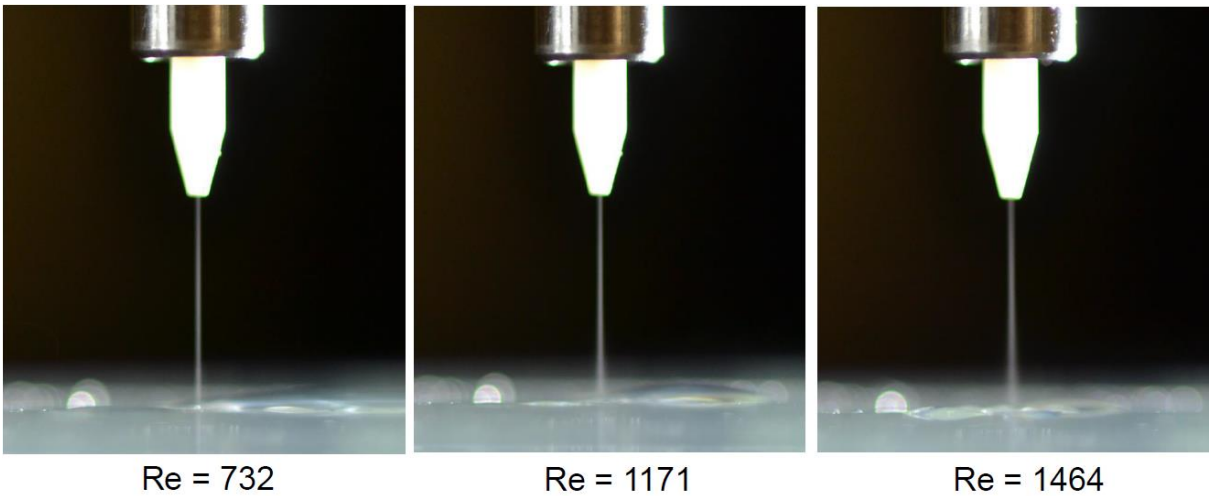

Fig. 5 As Fig. 4 but for $S = 6$ mm at $Re = 732$, 1171, and 1464 (for $Q = 150$, 240, and 300 sccm).

For impinging jets with standoff $S$ = 10 mm (~ 33x$D$), Figure 4 shows that the level of turbulent diffusion decreases with reducing the value of $Re$ from 1171 ($Q$ = 240 sccm) to 878 ($Q$ = 180 sccm). The impinging mist jet at $Re$ = 586 appears completely laminar, because its corresponding free-jet breakdown length (in Fig. 2) is greater than 10 mm. Reducing the standoff to $S$ = 6 mm (= 20x$D$) seems to further suppress turbulence development, as shown in Fig. 5 where even at Re = 1464 there is no obvious sign of turbulent diffusion, although the mist stream appears to widen slightly toward the substrate. According to Fig. 2 at $Re$ = 1171, a sign of turbulent diffusion as an abrupt widening of the mist stream appears beyond 5 mm from the nozzle tip. The measured breakdown length of free jet for Re = 1464 is 13 (in units of $D$, corresponding to 4 mm). Yet turbulent diffusion disappears even with an impingement wall located at $S$ = 6 mm, more than 4 mm away. Hence, the presence of impingement wall tends to interrupt turbulence development otherwise observable in the free-jet flow. But the wall effect diminishes with increasing standoff, as intuitively expected.

3.2 Cases of $D$ = 0.15 mm

With a smaller nozzle of $D$ = 0.15 mm (which is usually used for Aerosol Jet® fine feature printing), the jet velocity for the same $Re$ increases by a factor of two from that with $D$ = 0.3 mm. Figure 6 shows that at $Re$ = 390 the mist flow remains laminar for > 15 mm, but that at $Re$ = 586 breaks down at a length of 48 (in units of $D$) and at $Re$ = 878 the breakdown length becomes 29. By comparison with the case for $Re$ = 586 in Fig. 2, the breakdown length seems to increase from 40 to 48 when $D$ is reduced from 0.3 to 0.15 mm, as expected from stronger advection effect with higher jet velocity.

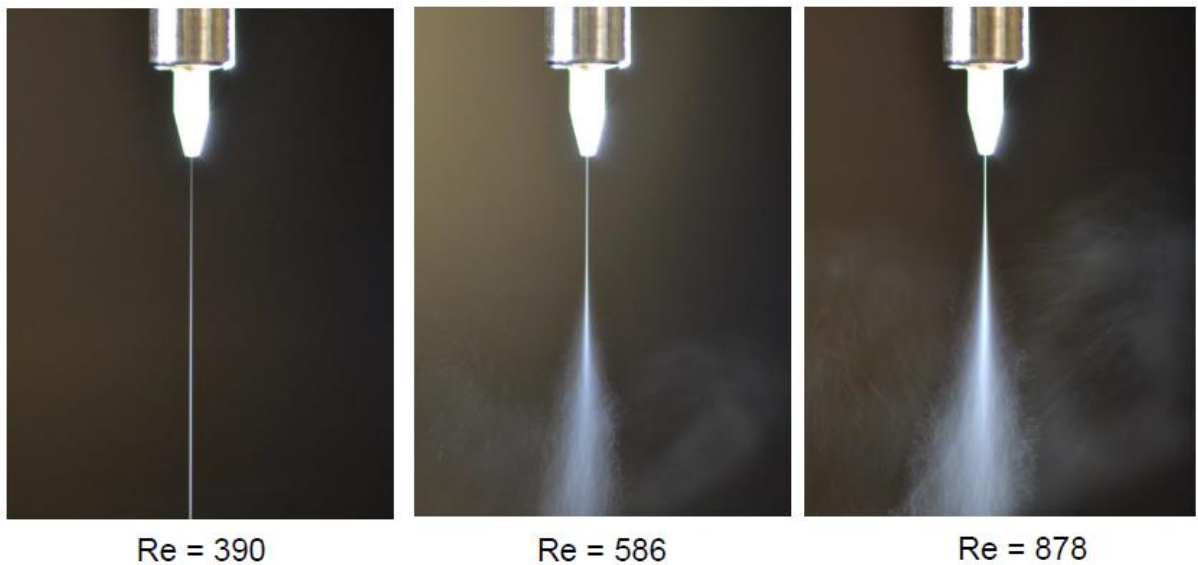

Fig. 6 Free-jet mist flows of $D$ = 0.15 mm at Re = 390, 586, and 878 (for $Q$ = 40, 60, and 90 sccm) with sheath-to-mist ratio $Y$ = 1:1. The length of the white ceramic nozzle tip outside the stainless steel holder is 4 mm, which can effectively serve as a scale reference.

For impinging jets, Figure 7 illustrates that for $S$ = 6 mm the mist stream starts to widen around 4 mm at $Re$ = 878, apparently consistent with a breakdown length of 30 indicated by the free jet in Fig. 6. The mist stream at $Re$ = 1171 breaks down at a short length. At Re = 586, the mist stream is well collimated because the corresponding free-jet flow has a breakdown length about 47 (i.e., 7 mm for $D$ = 0.15 mm). With the standoff shortened to $S$ = 4 mm (~27x$D$), mist flows appear collimated even up to $Re$ = 1464, without signs of turbulent diffusion (as illustrated in Fig. 8).

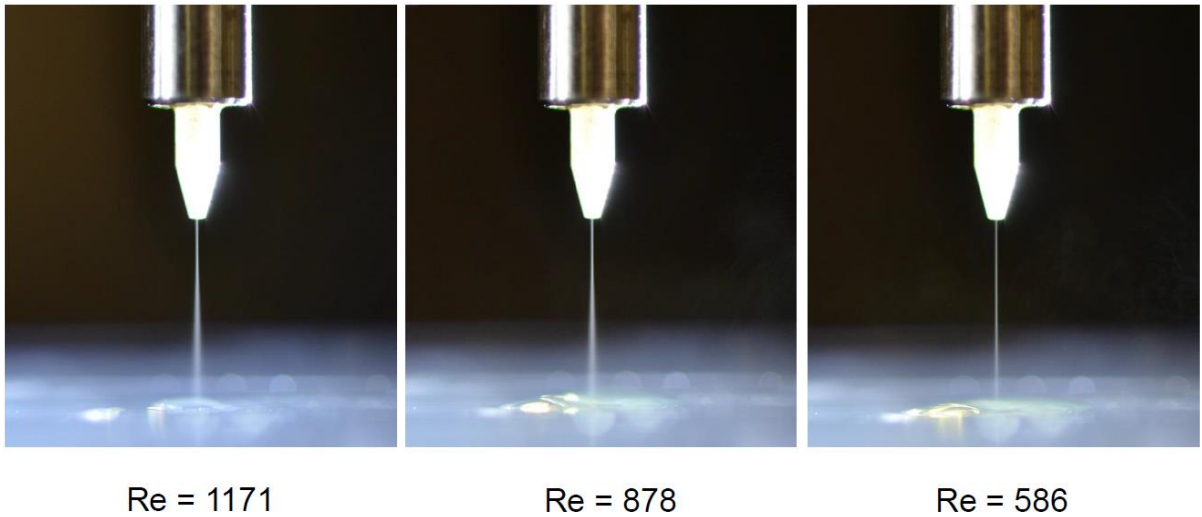

Re = 1171    Re = 878    Re = 586

Fig. 7 Impinging-jet mist flows of $D$ = 0.15 mm at Re = 1171, 878, and 586 (for $Q$ = 120, 90, and 60 sccm) with sheath-to-mist ratio $Y$ = 1:1, for $S$ = 6 mm.

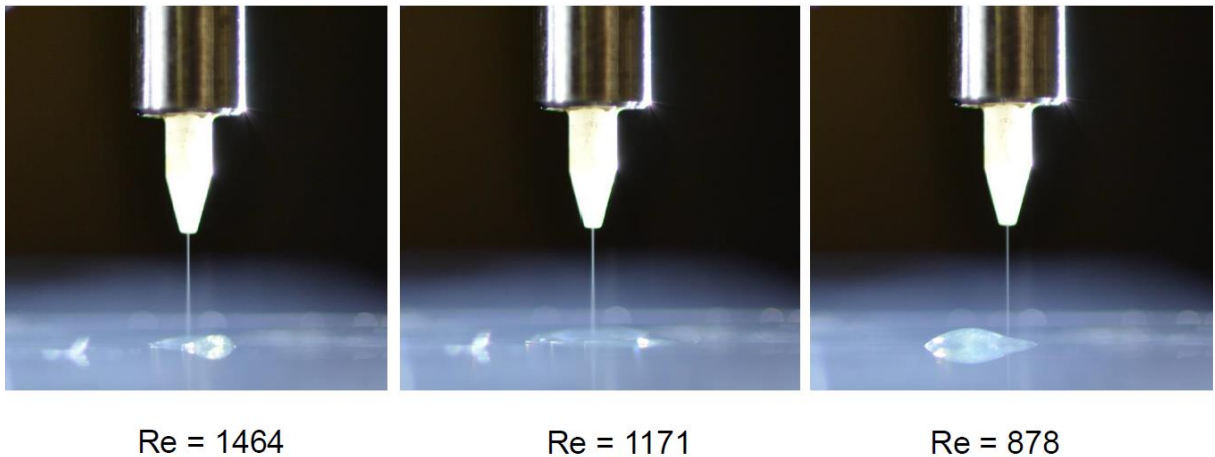

Re = 1464    Re = 1171    Re = 878

Fig. 8 As Fig. 7 but for $S$ = 4 mm at Re = 1464, 1171, and 878 (for $Q$ = 150, 120, and 90 sccm).

3.3 Cases of *D* = 1.0 mm

The nozzle of *D* = 1.0 mm is for printing high mass-output large features with Aerosol Jet® system. As a scale reference, the outer diameter of the nozzle at its tip is 2.75 mm. At *Re* = 1464 (for *Q* = 1000 sccm), Figure 9 shows the free jet has a breakdown length of about 8 (comparable to that of the visualization image for *Re* = 1305 with *D* = 4.0 mm by Kwon and Seo, 2005), and there is considerable turbulent diffusion of mist flow after 10 mm downstream for the case with standoff *S* = 20 mm. As *S* is reduced to about 12 mm (which is still > 8 or 10 mm), the turbulent diffusion diminishes with the mist stream becoming fairly collimated. The same visualization image as that for *Re* = 1464 at *S* = 12 mm is also seen with *Re* up to 4392 with *Q* = 3000 sccm, surprisingly. Again, a close enough impingement wall seems to effectively suppress turbulence development.

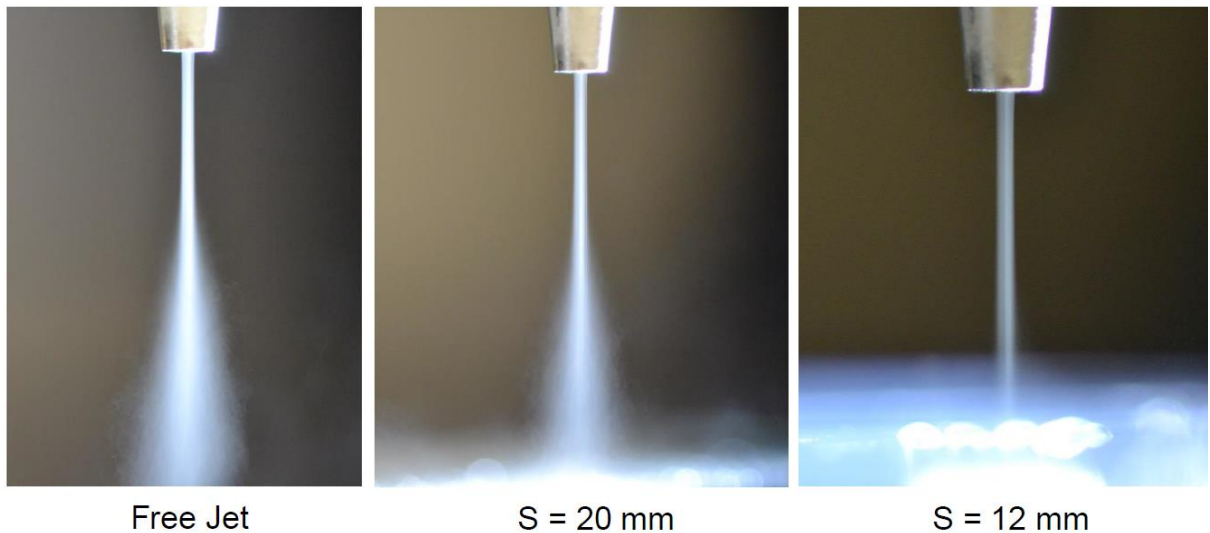

Fig. 9   Mist flows of *D* = 1.0 mm at Re = 1464 (for *Q* = 1000 sccm) with sheath-to-mist ratio *Y* = 1:1 for different standoff distances. The outer diameter of the nozzle tip is 2.75 mm, which can effectively serve as a scale reference.

On the other hand, significant mist stream widening due to viscous diffusion of momentum (instead of turbulent diffusion) can occur at small *Re* values, as illustrated by Fig. 10 for *Re* = 293. Such a viscous diffusion effect can also be suppressed by bringing the impingement wall closer to the nozzle (as shown in Fig. 10 with *S* = 12 mm). The viscous diffusion effect seems to be much more salient for large nozzles (such as that of *D* = 1.0 mm) than those small ones for fine-feature printing.

Because the value of *Re* is calculated as $\rho U D / \mu$, increasing *D* for a given value of *Re* means a proportional reduction of jet velocity *U*. Lowering the jet velocity *U* corresponds to reduced intensity of mist advection in the jet flow direction, allowing the lateral viscous diffusion effect to become more noticeable. This might be easily seen by comparing the image of $Re = 293$ in Fig. 2 for $D = 0.3$ mm, where widening of the mist stream does not appear as significant at a comparable distance of 6 mm ($20 \times D$) from nozzle tip presumably due to stronger mist advection effect with higher jet velocity.

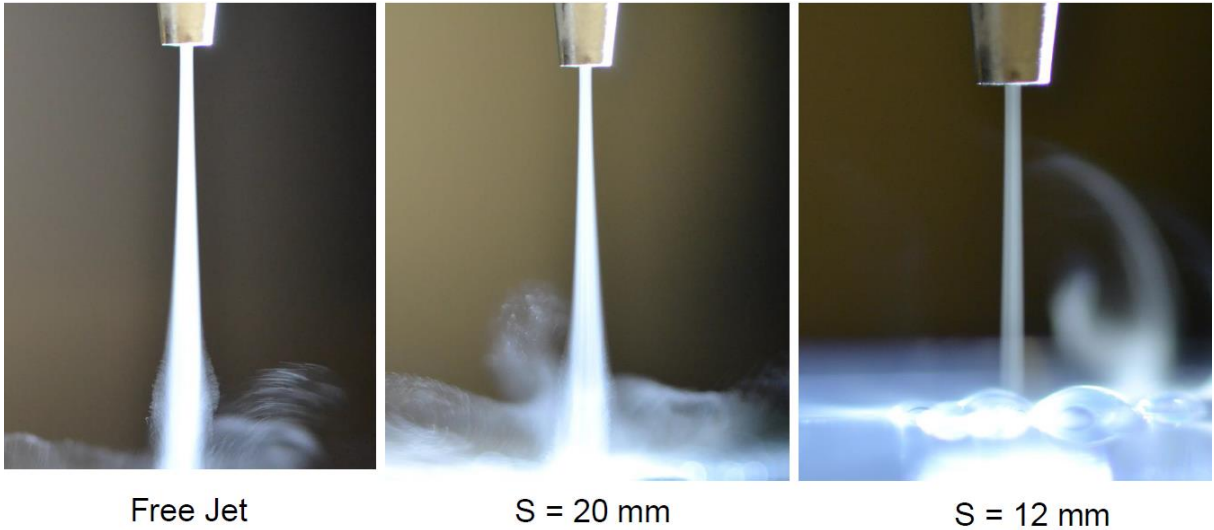

Fig. 10  As Fig. 9 but at Re = 293 (for $Q = 200$ sccm).

## 4 Concluding Remarks

With the Aerosol Jet® direct-write system, the flow pattern of jet core can be visualized with the scattered light from water microdroplets in the mist stream. The adjustable sheath-to-mist flow ratio offers a "peeling the onion" capability for visualizing jet flow structures. The images of mist flow visualization for free jets exhibit a laminar potential core extending some distance downstream from the nozzle, before developing turbulent eddies signaled as rapid widening of mist stream due to turbulent diffusion. The (dimensionless) breakdown length $L/D$—defined as the distance from nozzle tip to the inception of turbulent diffusion—decreases with increasing jet Reynolds number $Re$ as qualitatively consistent with previous authors (e.g., Becker and Massaro, 1968; Kwon and Seo, 2005; Gau et al., 2009). Although the exact value of breakdown length is difficult to measure with the available instruments used here, the estimated values based on the images of mist flow visualization with sheath-to-mist ratio $Y = 1:1$ can be fitted to an empirical formula

$$L/D = A / Re^B , \qquad \text{for } 500 < Re < 1500 , \tag{2}$$

with $A$ and $B$ = $1.5 \times 10^5$ and 1.25, $8.0 \times 10^4$ and 1.20, $2.0 \times 10^3$ and 0.75 respectively for $D$ = 0.15, 0.3, 1 mm, in a range of $Re$ (= $1.464 \, Q / D$ with $Q$ in units of sccm and $D$ in mm) especially relevant to Aerosol Jet® printing. It should be noted that the value of apparent free-jet breakdown length can also vary according to sheath-to-mist ratio $Y$ (as illustrate in Fig. 3).

Equation (2) retains similar mathematical form as that presented by Becker and Massaro (1968) for round gas jets of $D = 6.35$ mm with $Re > 1450$, as well as that by Gau et al., (2009) for 2D gas jets of width $< 0.2$ mm with $Re < 320$, but with different values of fitting parameters $A$ and $B$. Such differences among fitting formulas from different authors are likely due to limited data covering different ranges of $Re$ which is still open for future research. Nevertheless, a formula like (2) can provide a useful guideline for determining the maximum gas flow rate $Q$ for having a laminar mist stream with given $D$ and $S$ in developing Aerosol Jet® printing process recipes. For example, a laminar mist stream is expected with $Q < 120$ sccm ($Re < 586$) for a nozzle of $D = 0.3$ mm at standoff $S = 12$ mm ($L / D = 40$).

In discussion of impinging jets, the critical value of $Re$ for transition to turbulence has not been clearly defined. The general description in the literature (e.g, Zuckerman and Lior, 2006) suggested laminar jet flow for $Re < 1000$ and transition to turbulence between $Re = 1000$ and 3000. Most traditional applications with impinging jets have been associated with heat-mass transfer enhancement, which motivated research works reported in the literature to have often focused on turbulent jets for $Re > 4000$ to yield much larger values of Nusselt number (e.g., Popiel and Trass, 1991).

However, the images of mist flow visualization in the present work show that the mist flow carried by impinging jets (with $D = 1.0$ mm or less) may remain laminar for even $Re > 4000$ when the impingement wall is placed $< 12$x$D$ from the jet nozzle. With smaller nozzles of $D = 0.3$ mm and less, the impinging mist jets appear to be laminar and well collimated for $Re > 1000$ with standoff $S \sim 20$x$D$. Although the mist flow visualized here may remain laminar for longer distance than the entire jet including the surrounding sheath gas, the actual breakdown distance, if considered to be located downstream beyond which the potential core diminishes, should be revealed with reasonable accuracy in the visualized mist flow pattern shown here.

**Acknowledgements** The author would like to thank Dr. Kurt Christenson and Dr. Mike Renn for support and helpful discussions.